\documentclass[manuscript,nonacm]{acmart}

\AtBeginDocument{%
  }

\setcopyright{none}
\copyrightyear{2025}

\acmConference[Conference acronym 'XX]{Make sure to enter the correct
  conference title from your rights confirmation email}{June 03--05,
  2018}{Woodstock, NY}
\acmISBN{978-1-4503-XXXX-X/2018/06}

\begin{document}

\title{OnomaCompass: A Texture Exploration Interface that Shuttles between Words and Images}



\author{Miki Okamura}
\affiliation{%
  \institution{University of Tsukuba}
  \country{Japan}}
\email{mikio_kamura@digitalnature.slis.tsukuba.ac.jp}

\author{Shuhey Koyama}
\affiliation{%
  \institution{University of Tsukuba}
  \country{Japan}}
\email{shuhey@digitalnature.slis.tsukuba.ac.jp}

\author{Li Jingjing}
\affiliation{%
  \institution{University of Tsukuba}
  \country{Japan}}
\email{li@digitalnature.slis.tsukuba.ac.jp}

\author{Yoichi Ochiai}
\affiliation{%
  \institution{University of Tsukuba}
  \country{Japan}}
\email{wizard@slis.tsukuba.ac.jp}

\renewcommand{\shortauthors}{Okamura et al.}

\begin{abstract}
Humans can finely perceive material textures, yet articulating such somatic impressions in words is a cognitive bottleneck in design ideation. We present OnomaCompass, a web-based exploration system that links sound-symbolic onomatopoeia and visual texture representations to support early-stage material discovery. Instead of requiring users to craft precise prompts for generative AI, OnomaCompass provides two coordinated latent-space maps—one for texture images and one for onomatopoeic terms—built from an authored dataset of invented onomatopoeia and corresponding textures generated via Stable Diffusion. Users can navigate both spaces, trigger cross-modal highlighting, curate findings in a gallery, and preview textures applied to objects via an image-editing model. The system also supports video interpolation between selected textures and re-embedding of extracted frames to form an emergent exploration loop. We conducted a within-subjects study with 11 participants comparing OnomaCompass to a prompt-based image-generation workflow using Gemini 2.5 Flash Image (“Nano Banana”). OnomaCompass significantly reduced workload (NASA-TLX overall, mental demand, effort, and frustration; $p < .05$) and increased hedonic user experience (UEQ), while usability (SUS) favored the baseline. Qualitative findings indicate that OnomaCompass helps users externalize vague sensory expectations and promotes serendipitous discovery, but also reveals interaction challenges in spatial navigation. Overall, leveraging sound symbolism as a lightweight cue offers a complementary approach to Kansei-driven material ideation beyond prompt-centric generation.
\end{abstract}

\ccsdesc[500]{Human-centered computing~Interactive systems and tools}
\ccsdesc[500]{Human-centered computing~Interaction paradigms}
\ccsdesc[300]{Human-centered computing~HCI design and evaluation methods}
\ccsdesc[300]{Human-centered computing~Interaction techniques}

\keywords{Perceptual Augmentation, Creativity Support Tools, Sound Symbolism, Latent Space Exploration, Onomatopoeia}

\begin{teaserfigure}
  \includegraphics[width=\textwidth]{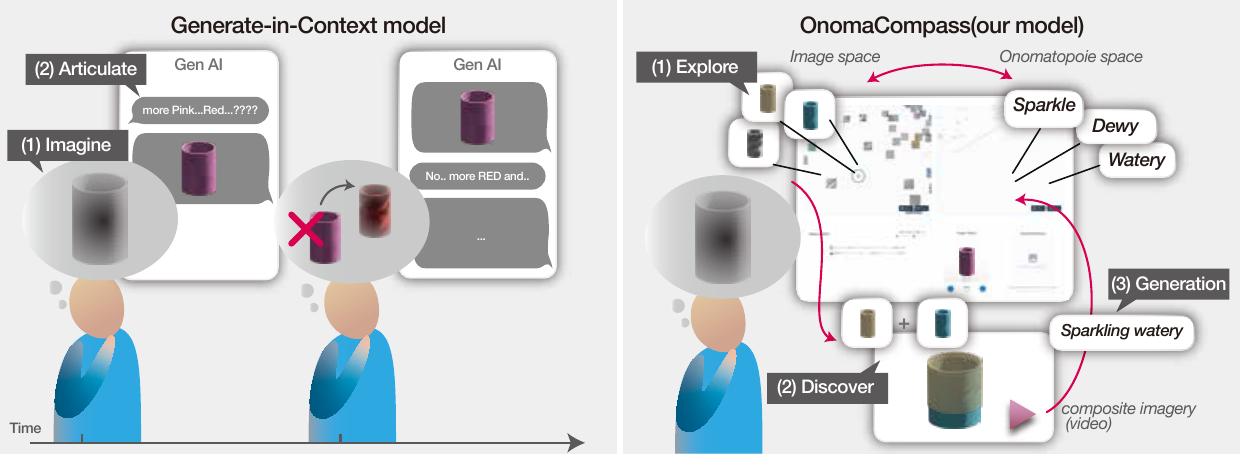}
  \caption{Conceptual comparison of prompt-centric generation and OnomaCompass: rather than iteratively verbalizing intent (left), users shuttle between linked maps of textures and onomatopoeia (right), using cross-modal cues to navigate materiality, apply textures to objects, and extend exploration through video interpolation and frame re-embedding.}
  \label{fig:teaser}
\end{teaserfigure}


\maketitle

\section{Introduction}
\subsection{A new frontier of creativity: Designing “materiality”}
Advancements in digital technologies have expanded the agency of creation from experts to the general public, ushering in an era in which anyone can become a creator \cite{Shneiderman2007}. Within this broader movement, interest in creativity has begun to shift from designing mere form to designing materiality, which determines a product’s impression and tactility. The scope of the term “materiality” as used in this study is defined explicitly. In specialized domains such as 3DCG and industrial design, “texture” often denotes a two-dimensional image applied to a surface, whereas “material” frequently refers to a set of optical parameters. In contrast, to align with the everyday sensemaking of novices without specialized knowledge, this study adopts a broader conception of materiality. Specifically, “materiality” refers to the overall visual and haptic impression perceived by humans, integrating both the intrinsic optical qualities of a substance, namely the material appearance, and the surface geometry, which includes fine irregularities and motion, such as an aggregation of feathers or the undulations of gels. For this emerging design domain, especially for novices who are not specialists, methods that enable navigation from vague, sensorial images to concrete design concepts have yet to be established. This study focuses on ideation, the most upstream stage of the creative process.

\subsection{The “verbalization barrier” in generative AI}
Recent text-to-image systems have offered a powerful response to this challenge. However, the prevailing interaction paradigm intrinsically depends on the user’s capacity for verbalization \cite{Furnas1987}. Contemporary creative dialogue with AI typically proceeds in turns: users must translate their internal images into discrete words as “prompts,” and the AI interprets these to return “images” \cite{Liu2022}. While effective when users possess explicit linguistic vocabularies, this process becomes a bottleneck when handling sensorial images that resist verbalization. This verbalization barrier has been noted to reduce opportunities for serendipitous discovery and to constrain creative potential.

\subsection{A “shuttling” between words and images: A dialogue that bridges two worlds}
To address this challenge, a new interaction paradigm is proposed that transforms human-AI dialogue from a one-way, instruction-response model into a dialogic process that shuttles bidirectionally between the worlds of words (concepts) and images (perception). This idea draws on cognitive processes in which thought deepens by continually moving between concrete perception and abstract conceptualization. The implemented system, “OnomaCompass,” constructs two distinct representational spaces: an image space that captures visual features \cite{Karras2019} and a language space that captures semantic relatedness. By interactively shuttling between these two spaces, users can progressively concretize their ambiguous images through both visual discovery and conceptual understanding.

\subsection{Onomatopoeia: A bridge to the sensorial world and an AI’s vision}
As the bridge for navigating between these spaces, the study focuses on Japanese onomatopoeia, with particular emphasis on mimetics. In broader linguistic contexts, these are known as ideophones \cite{Akita2016}. The approach adopted here uses onomatopoeia as starting points for instructing AI-based image generation. The resulting images include not only close-ups of material surfaces but also concrete motifs that symbolically express specific materialities (for example, a sweet potato for “hokuhoku”). This observation reflects an intriguing property: when visualizing linguistic concepts of materiality, AI frequently responds by depicting prototypical objects that possess the target material quality. This study regards such concrete motifs as effective visual expressions for conveying materiality and includes them in the dataset.

\subsection{Proposal and contributions}
Based on these ideas, this paper presents “OnomaCompass”, a new interface that supports sensorial materiality design through exploration of continuous latent spaces. Unlike prior latent space exploration interfaces designed primarily for expert analysis and control, this system targets novices engaged in creative activity and addresses their sensorial needs that are difficult to verbalize. To achieve this, a dialog model was implemented that connects and enables shuttling between latent spaces of different modalities, a capability made feasible by recent advances in multimodal AI. The contributions are threefold. First, a proof of concept is provided for a new HCI paradigm that shifts from discrete linguistic instructions to continuous spatial shuttling in human-AI creative dialogue. Second, this paradigm is instantiated in a concrete system that uses onomatopoeia as a bridge to connect two latent spaces and to perform dynamic video interpolation. Third, semi-structured interviews with eleven participants show that the proposed approach offers qualitatively distinct value—namely, the joy of serendipitous discovery and relief from the pressure to verbalize—especially during the divergent thinking stage for novices. The remainder of the paper reviews related work in Section 2, details the methodology in Section 3, reports the evaluation results in Section 4, discusses the findings in Section 5, and concludes with future directions in Section 6.

\section{Related Work}
This study lies at the intersection of three areas: human-AI creative dialogue, exploration of latent spaces, and the cognitive science of onomatopoeia. This section reviews prior work in these areas to clarify the problem addressed and the originality of the approach.

\subsection{Discreteness and continuity in creative dialogue}
The core of the proposed interaction is a shift from discreteness to continuity.

\subsubsection{Challenges of text prompts: Semantic gap and discrete trial-and-error}
Interaction with current image-generation AI is primarily mediated by text prompts. While powerful, a well-known semantic gap exists between the user’s continuous, sensorial images and the discrete linguistic expressions required by the system, a gap that has long been recognized in information retrieval \cite{Furnas1987}. Users are often forced to bridge this gap by iteratively revising prompts and regenerating outputs \cite{Liu2022}. This process can impose high cognitive load, especially on novices without a well-defined vocabulary, and can disrupt the flow of creative thinking.

To address these challenges, recent systems have introduced interactive support for prompt engineering using Large Language Models (LLMs) or attention visualization mechanisms \cite{brade2023promptify, wang2024promptcharm}. Large-scale prompt datasets have also been constructed to help understand how models react to different linguistic inputs \cite{wang2023diffusiondb}. Notably, PromptMap \cite{adamkiewicz2025promptmap} proposes an alternative interaction style that allows users to freely explore vast collections of synthetic prompts through a map-based interface with semantic zoom. Despite these advancements in text-based assistance, reliance on linguistic descriptions remains a barrier for purely sensory or abstract ideation.

\subsubsection{Efficacy of continuous interfaces}
HCI research demonstrates that continuous parameter manipulation, such as dynamic queries, significantly enhances exploratory tasks compared to discrete value entry \cite{Ahlberg1994}. However, merely increasing control dimensions without semantic organization exacerbates cognitive load \cite{dang2022ganslider}. Consequently, effective creative exploration requires interfaces that facilitate intuitive visual navigation within semantically structured latent spaces, rather than simply exposing raw parameters.

\subsection{Navigating the sensorial world: Exploration of latent spaces}
This study follows the line of research that develops interfaces for interactive exploration of latent spaces learned by AI \cite{Goodfellow2014, Karras2019}. 

\subsubsection{Foundations of Unified Representation}
Recent advancements have expanded latent spaces beyond single modalities. Unified representation spaces have been established to align multiple modalities—such as video, audio, depth, and thermal data—using language as a binding element \cite{girdhar2023imagebind, zhu2023languagebind, lyu2024unibind}. Such multimodal spaces allow for richer semantic alignment, enabling the combined analysis of distinct data types \cite{kwon2023latent}.

\subsubsection{Visualization and Manipulation Tools}
To navigate these high-dimensional spaces, various visualization tools have been developed. These include scalable visual analytics to map and compare semantic dimensions \cite{liu2019latent, wang2023wizmap}, as well as browsers for large image datasets or cultural collections via dimensionality reduction \cite{grotschla2024aeye, ohm2023collection}. In the video domain, projecting frames into navigable latent spaces has been shown to support brainstorming \cite{lin2024videomap}.

On the manipulation side, research has focused on discovering controllable directions within these spaces. Several methods identify interpretable directions (e.g., pose, age) in GAN latent spaces via PCA or linear subspaces \cite{harkonen2020ganspace, shen2020interfacegan}. More recently, text-guided manipulation techniques have enabled robust zero-shot editing using natural language \cite{kim2022diffusionclip, zhu2022one, kocasari2022stylemc, baykal2023clip}. 

\subsubsection{Ideation-Focused Interfaces}
While the aforementioned tools often target precise control or analysis, there is a growing interest in interfaces specifically for creative ideation. Recent systems facilitate the recombination of visual references and keywords for mood board creation \cite{wan2023gancollage, choi2024creativeconnect}, or bridge tacit user intent with generation by allowing navigation through high-level semantic concepts \cite{lee2025thematicplane, abuzuraiq2025designing}. However, interfaces grounded specifically in sensorial language to support novices in exploring material appearance remain underexplored.

\subsection{A bridge to sensation via onomatopoeia}
To navigate this uncharted territory, the study focuses on the sound symbolism of Japanese onomatopoeia, especially mimetics. Sound symbolism refers to a non-arbitrary (intuitively connected) relationship between the sound of a word and its meaning.

\subsubsection{Intuitive understanding and reduced cognitive load}
This intuitive link between sound and meaning \cite{Maurer2006} suggests that onomatopoeia impose low cognitive load and are easy to learn. The "sound symbolism bootstrapping hypothesis" \cite{imai2014sound} argues that this sensitivity is biologically endowed and plays a critical role in language acquisition and establishing lexical representations. Research indicates that mappings between sound and meaning facilitate word learning \cite{Imai2008}, and this intuitiveness can transcend language barriers \cite{Iwasaki2007}. These findings provide strong grounds for regarding onomatopoeia as effective, intuitive interface elements for novices.

\subsubsection{Strong ties to materiality}
Onomatopoeia is deeply linked to multisensory perception \cite{Spence2011}. In the perception of visual materiality, onomatopoeia have been shown to capture basic perceptual dimensions such as wetness, fluffiness, and smoothness \cite{hanada2016using}. Computational approaches have also attempted to utilize this link, such as computer vision systems that express texture using sound-symbolic words by mapping phonemes to texture categories \cite{yamagata2021computer}. Studies using functional MRI have shown that the right posterior STS is consistently activated when processing Japanese mimetics across different domains, such as motion and shape, reflecting their role as both linguistic and iconic symbols \cite{kanero2014sound}. These studies ground the scientific validity of applying onomatopoeia to the task of exploring material appearance.

\subsection{Generative AI and the Creation of Materiality} This study also situates itself within the context of AI-driven materiality generation. Recent systems such as TactStyle \cite{faruqi2025tactstyle} generate haptic surface properties from image prompts for digital fabrication. While TactStyle focuses on the faithful reproduction of specific materialities, our approach targets the upstream ideation phase. By leveraging onomatopoeia as sensorial cues, our method enables users to navigate the conceptual space between diverse textures. Through generative interpolation, the proposed system facilitates the discovery of novel material qualities that emerge within these transitional states.

\section{Methodology}
To assess the effectiveness of the proposed system, we implemented a prototype and evaluated it in a within-subjects mixed-methods study combining standardized questionnaires and semi-structured interviews.

\begin{figure}[h!]
\centering
\includegraphics[width=0.95\textwidth]{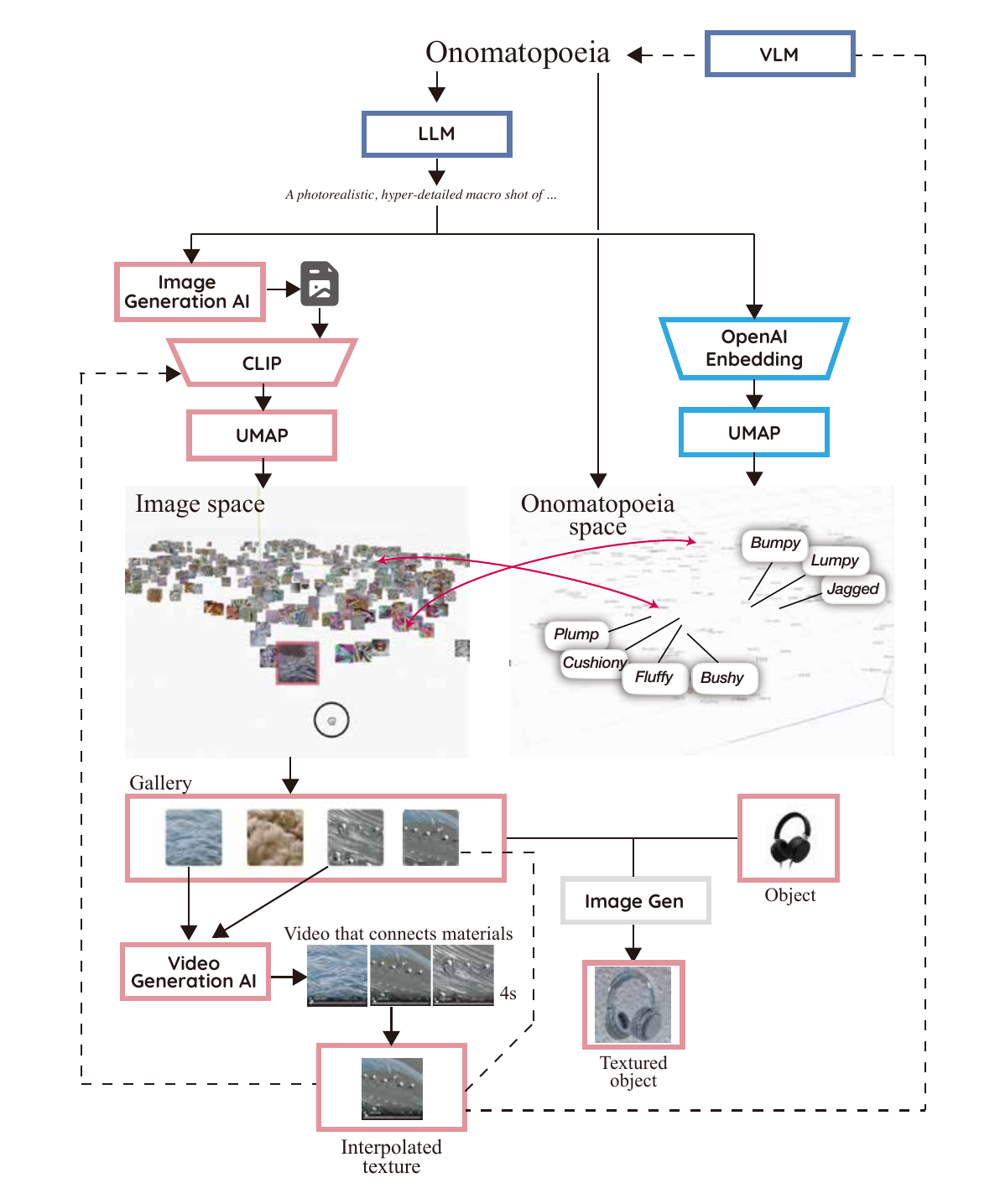}
\caption{\textbf{OnomaCompass system overview:} dual visualization of image and language latent spaces with cross-modal highlighting, a gallery for curation and texture application to objects, and a dynamic video interpolation and replotting pipeline that feeds newly generated frames back into both spaces.}
\label{fig:system}
\end{figure}

\subsection{System Design and Implementation}
This section describes the specific design and implementation of the proposed “OnomaCompass” system. To realize the theoretical framework presented in Section 1, a web-based system was built to enable users to shuttle interactively between visual and linguistic latent spaces.

\subsubsection{Design goals}
To instantiate the concept of shuttling between words and images, four design goals were established.

\paragraph{DG1: Reduce cognitive load}
Build a system that circumvents the explicit verbalization demanded by conventional image-generation AI and enables creative exploration through intuitive visual and spatial manipulation. Particular emphasis was placed on allowing novices without specialized knowledge to begin exploring materiality without mastering complex prompt engineering.

\paragraph{DG2: Promote serendipity}
Actively induce serendipitous discoveries that are not constrained by users’ existing knowledge and vocabulary. This approach fundamentally differs from conventional search methods based on predefined queries. By visualizing semantic proximity through spatial layout and presenting cross-modal correspondences, the system provides an environment in which users can uncover unexpected relationships.

\paragraph{DG3: Enable a continuous exploration experience}
Provide a continuous exploration experience through fluid navigation in a continuous spatial layout (2D maps visualized in a navigable 3D view), rather than discrete trial-and-error. The interaction design aimed to allow natural shuttling between visual and linguistic representations without disrupting the flow of thought.

\paragraph{DG4: Construct an emergent loop}
Realize a system in which users’ acts of exploration themselves generate new targets of exploration. In particular, dynamic content generation and replotting mechanisms were developed to extend the scope of exploration beyond a fixed dataset.

\subsubsection{Key features and interface}
To achieve the above design goals, the system implements the following five core features.

\begin{figure}[h!]
\centering
\includegraphics[width=0.9\textwidth]{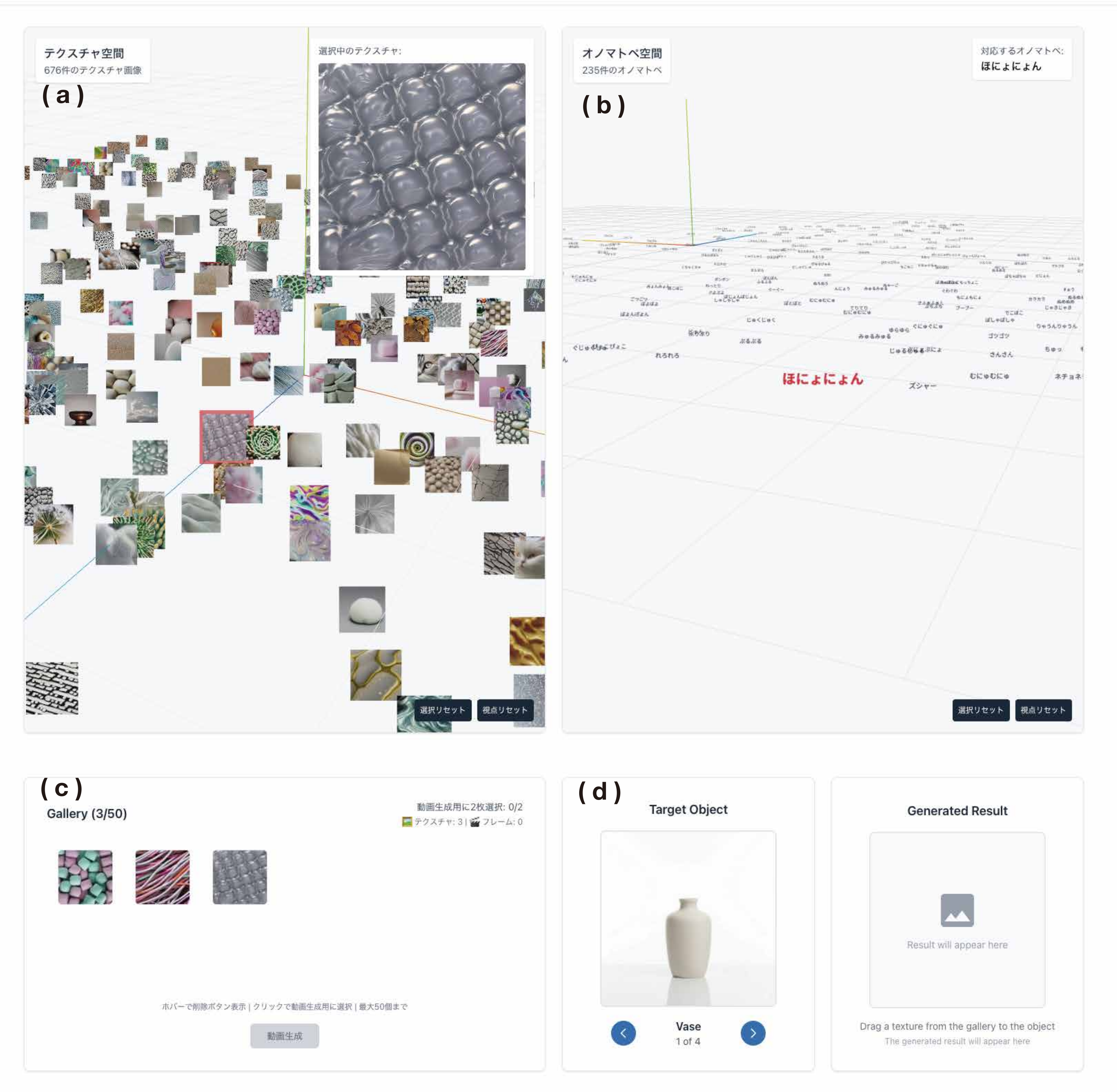}
\caption{\textbf{The user interface of the proposed system.} (a) Texture Space: A scatter plot of texture images arranged based on visual features. (b) Onomatopoeia Space: A scatter plot of onomatopoeic terms arranged by text-embedding similarity of the LLM-generated prompt descriptions (written to reflect sound-symbolic impressions; e.g., "Honyonyon"). (c) Gallery and Preview: An area for managing selected textures and simulating their application onto a target object (e.g., a vase).}
\label{fig:ui}
\end{figure}

\paragraph{Dual Latent Space Visualization}
At the core of the system is a parallel visualization of the image latent space and the language latent space (Figure~\ref{fig:ui}). Two coordinated maps are presented side by side. Each map is computed in 2D via UMAP and rendered in a 3D scene for navigation. The left image space contains 676 texture images, and the right language space contains 235 onomatopoeic terms, each placed at 2D coordinates (with a fixed $z$ value) in the 3D view.
The system centers on two latent spaces comprising onomatopoeia and their visual expressions. Because the aim is to support creative exploration of materiality unconstrained by generic object concepts, the dataset was intentionally constructed using invented onomatopoeia. Rather than limiting entries to familiar everyday terms, this choice was made to stimulate creativity by prompting imagination through sound symbolism.
First, for invented onomatopoeia collected by the author and from prior projects, a large language model (LLM) was used to generate image prompts in stages. This process involved identifying the material evoked by the onomatopoeia, defining its physical qualities, and producing an English rendering, culminating in prompts for Stable Diffusion. Stable Diffusion was selected not for state-of-the-art photorealism but to afford artistic, diverse interpretations of materiality.
Each space is rendered in the browser using WebGL via Three.js. Users can freely change viewpoints with mouse operations: drag to rotate, wheel to zoom, and right-drag to pan. Images are loaded as 64×64 px WebP thumbnails and displayed as sprites scaled by baseScale = 0.075 (approximately 4.8×4.8 equivalent) so they always face the camera during view changes, maximizing recognizability within the 2D maps visualized in a navigable 3D view.
\paragraph{Cross-Modal Highlighting}
As a core function for DG2 (serendipity), cross-modal highlighting was implemented. When an item is selected in one space, related items in the other space are emphasized based on precomputed cross-modal links defined by the dataset authoring pipeline (the shared material-description text used to generate textures for each invented term). In other words, cross-modal relations are not inferred online by similarity search; they are defined by the authoring pipeline that connects each invented onomatopoeia to its material description and the textures generated from it. Selecting an onomatopoeia highlights its associated texture images (1--3 images, depending on generation success) and shows them in a small preview panel, allowing users to choose among variations. Selecting a texture image highlights the single onomatopoeic term from which it was generated. This asymmetric mapping supports two complementary exploration patterns: “What images can this word evoke?” and “What word can describe this image?”
\paragraph{Gallery and Texture Application}
To save and utilize textures discovered during exploration, a gallery function was implemented. Saved images can be dragged and dropped onto presented object images (a vase and headphones) to generate previews of texture application.
For texture application, Gemini 2.5 Flash Image (image generation and editing) is used to compose the texture image—capturing both material appearance and surface geometry—with the object image to produce the applied-material image.
\paragraph{Dynamic Video Interpolation and Re-plotting}
To realize DG4 (an emergent loop), a dynamic video interpolation and replotting system was developed. Users can select two images in the gallery and generate a video expressing visual transitions between them.
Video generation uses Luma AI Ray 1.6 to perform one-way interpolation (A→B) with an emphasis on vital, fluid “material behavior,” rather than simple morphing. Concretely, a long prompt is used to produce an ultra-high-resolution, photorealistic macro shot from a fixed camera position, including localized organic and fantastical phenomena (fluid transitions, emergent lifeforms, phase transitions of matter, and so forth).
Furthermore, arbitrary frames can be extracted from the generated video and repositioned into the existing latent spaces through a four-stage AI pipeline:
Step 1: Visual feature extraction Using the VLM capability of Gemini 2.0 Flash (gemini-2.0-flash-exp), the extracted frame is analyzed for visual features and corresponding onomatopoeia are generated directly.
Step 2: Concept description generation Given the onomatopoeia from Step 1, the OpenAI o1 model generates a detailed concept description (description), which provides contextual semantic information for the subsequent embedding process.
Step 3: Feature embeddings Image feature: The frame is embedded into a 512-dimensional vector using CLIP ViT-B/32. Text feature: The concept description from Step 2 is embedded into a 1536-dimensional vector using OpenAI text-embedding-3-small.
Step 4: Projection into latent spaces Using a pre-trained UMAP model, the features from Step 3 are projected into the existing two-dimensional coordinate spaces. Coordinates corresponding to the texture space and the onomatopoeia space are computed, enabling placement of new data without retraining.
Replotted dynamic points are distinguished in orange, realizing an emergent loop in which users’ exploratory actions generate new targets for further exploration.

\subsubsection{Implementation details}
\paragraph{System architecture}
OnomaCompass adopts a frontend-centric architecture to preserve responsiveness during continuous exploration (DG3). Embeddings and map coordinates are precomputed and delivered to the client so that most interactions can be handled in the browser without repeated server round-trips. The backend is implemented with Python 3.12 and FastAPI 0.104, and the frontend is implemented with React 19.1.1 and TypeScript 5.8.3. The visualization is rendered with Three.js r162 via React Three Fiber.

\paragraph{Dataset construction and embedding pipeline}
We curated 235 invented onomatopoeic terms. For each term, we generated a Stable Diffusion prompt in stages using OpenAI o1 (onomatopoeia $\rightarrow$ material identification $\rightarrow$ definition of physical qualities $\rightarrow$ English rendering $\rightarrow$ Stable Diffusion prompt), and then generated three texture images per term. Due to occasional generation failures, some terms have fewer than three associated textures, resulting in 676 images in total. This same description text also defines the cross-modal links in the interface, connecting each invented term to the set of textures generated from it for deterministic cross-modal highlighting. To construct the two coordinated maps, we computed image and language embeddings and projected them into two-dimensional coordinates with UMAP (n\_neighbors=15, min\_dist=0.5, metric=cosine). Image embeddings were computed as 512-dimensional vectors using CLIP ViT-B/32 \cite{Radford2021}. Language embeddings were computed as 1536-dimensional vectors using text-embedding-3-small applied to the generated prompt text (i.e., the English description used to generate textures with Stable Diffusion). Importantly, this language representation serves as a proxy for the intended material concept associated with each invented term; it does not explicitly model phonetic sound symbolism. In prompt generation, we explicitly instructed the LLM to preserve and articulate the intended sound-symbolic (mimetic) impression of each term when producing the English description. This shared prompt text serves as a common representation that links modalities in our prototype: it is used (i) to generate texture images with Stable Diffusion, (ii) to compute the language-map layout via text embeddings, and (iii) to define cross-modal links for deterministic highlighting.

\paragraph{Visualization and interaction}
The resulting 2D coordinates are visualized as two side-by-side maps in a navigable 3D view (with a fixed $z$ value) to support camera-based exploration. Texture images are rendered as sprites to maintain performance, and thumbnails (64$\times$64 px WebP) are loaded progressively with full-resolution images loaded on demand. Users can rotate, zoom, and pan using OrbitControls, and the camera can auto-focus on selected items to support rapid shuttling between the texture and onomatopoeia maps.

\subsection{Evaluation study}
We do not evaluate objective “best texture” outcomes in this study. Instead, we focus on how interfaces shape early-stage ideation—workload, exploratory behavior, and the externalization of sensory expectations—where goals may still be under formation. A mixed-methods evaluation centered on semi-structured interviews was designed to elucidate how the proposed system affects users’ creative thinking.

\subsubsection{Participants}
Eleven participants (undergraduate and graduate students) who were native Japanese speakers and agreed with the study’s background and purpose took part in the experiment.

\subsubsection{Task and procedure}
Participants were informed that the study envisioned “a future in which novices can create artifacts with their desired materiality.” They used both OnomaCompass (OC) and the baseline (NB) to explore and generate materiality for two targets: headphones and a vase.
Each condition lasted 15 minutes: 7 minutes for one target and 7 minutes for the other (plus transitions). The order of targets within each condition was alternated.
After each condition, participants completed questionnaires.
\paragraph{Baseline condition: Nano Banana}
Participants were provided with a base image of the target object (a vase or headphones) and were allowed to (i) start from the provided object image and iteratively edit it, or (ii) generate texture images from scratch and then generate an object image incorporating the texture. Additional reference images were not explicitly allowed. Participants could create multiple new chat threads and use the chat for clarification and ideation if desired, and no custom system prompt was set.
\paragraph{OnomaCompass condition}
All system features described in Section 3.1.2 (shuttling between spaces, gallery, texture application, video interpolation and replotting) were explained, followed by one minute of familiarization.
A within-subjects design was employed. To counterbalance order effects, six participants used OnomaCompass first and five used the image-generation AI first.

\subsubsection{Data collection and analysis}
After completing all tasks, semi-structured interviews were conducted. The interviews asked about: (i) overall impressions, (ii) strengths and weaknesses of each system, (iii) differences in materiality exploration processes, (iv) moments of surprise or interest, (v) use across different scenarios (rapid exploration versus ideation), (vi) likelihood of recommending to a friend, and (vii) areas for improvement in OnomaCompass. The verbal data were coded using qualitative thematic analysis, and emergent themes were organized and analyzed.

\section{Result}
\subsection{Quantitative evaluation}
This section reports the results of the UEQ, NASA-TLX, SUS, and an original questionnaire comparing OnomaCompass (OC) and Nano Banana (NB). The experiment used a within-subjects design ($N=11$). Participants' ratings were aggregated by condition for comparison. Normality was verified using the Shapiro–Wilk test. Paired t-tests were used for normally distributed data, and the Wilcoxon signed-rank test was used for data that did not satisfy normality. The significance level was set at $\alpha=0.05$. Reported p-values are two-tailed and uncorrected. Effect sizes are reported using Hedges' $g$ (corrected for small sample sizes) for t-tests and $r$ for Wilcoxon tests. Raw (unweighted) averages were used for NASA-TLX, and each item was interpreted such that a higher score indicates a higher workload.

\subsubsection{Core Value: NASA-TLX}
The overall workload (Raw average) was 34.46 ($SD=24.23$) for OC and 45.55 ($SD=25.56$) for NB. The OC condition was significantly lower [$t(10)=-2.686, p=0.023, g=-0.74$]. Consistent differences were also observed in the subscales. Specifically, Mental demand [38.36 vs. 58.00, $t(10)=-2.919, p=0.015, g=-0.80$], Effort [25.09 vs. 49.27, $t(10)=-2.269, p=0.047, g=-0.63$], and Frustration [23.46 vs. 56.82, $t(10)=-4.420, p=0.001, g=-1.22$] were all significantly lower in the OC condition. No significant difference was observed for Physical demand [$t(10)=1.518, p=0.160$]. Similarly, no significant differences were found for Temporal demand [$t(10)=0.403, p=0.695$] and Performance [$t(10)=-1.514, p=0.161$]. These results suggest that the proposed system significantly reduces cognitive resources (Mental, Effort) and emotional load (Frustration). Meanwhile, aspects related to physical operation and time pressure appear comparable between conditions. This finding aligns with the qualitative observations in Section~\ref{sec:qualitative} ("low vocabulary burden" and "OC is fun").

\subsubsection{UEQ: Superiority of Hedonic Value}
Regarding the composite indicators of the UEQ (-3 to +3), Hedonic Quality was significantly higher for OC [2.000 vs. -0.091, $t(10)=4.416, p=0.001, g=1.22$]. Overall Quality was also significantly higher for OC [1.578 vs. 0.694, $t(10)=2.847, p=0.017, g=0.78$]. Conversely, no significant difference was observed for Pragmatic Quality [1.159 vs. 1.477, Wilcoxon: $Z=-0.533, p=0.618, r=0.11$]. In other words, the proposed system significantly enhances hedonic aspects, such as enjoyment and novelty, and overall experience quality without compromising the level of practicality.

\subsubsection{SUS: Trade-off in General Usability}
SUS scores were significantly higher for NB [75.68 ($SD=12.25$) vs. 67.73 ($SD=17.69$), $t(10)=-2.316, p=0.043, g=-0.64$]. Conventional chat-based image generation appears superior in terms of general usability. This result can be interpreted as the influence of learning costs associated with operations specific to this prototype, such as 3D spatial navigation.

\subsubsection{Original Questionnaire: Divergence Support and Serendipity}
Among the 12 items rated on a 1–7 scale, significant differences were found in the following five items (uncorrected, exploratory analysis). Able to explore variations of diverse textures (Q3): OC > NB [6.73 vs. 4.00, $t(10)=4.404, p=0.001, g=1.21$]Got stuck not knowing what to do next (Reverse item, Q6): NB > OC [4.00 vs. 1.82, Wilcoxon: $Z=-2.934, p=0.002, r=0.63$]Mental load was low, and fatigue was minimal (Q7): OC > NB [5.46 vs. 3.73, $t(10)=3.012, p=0.013, g=0.83$]The process was creative and fun rather than technical (Q8): OC > NB [6.27 vs. 4.82, Wilcoxon: $Z=-2.667, p=0.008, r=0.57$]There were incidental discoveries that were better than expected (Q10): OC > NB [6.64 vs. 4.55, $t(10)=3.348, p=0.007, g=0.92$]Other items (e.g., sense of task accomplishment, clarification of vision, intention to reuse) did not reach statistical significance. However, many showed a trend favoring OC with medium to large effect sizes. Overall, both standard scales (NASA-TLX and UEQ) and the original items confirm that the proposed system has strengths in "avoidance of getting stuck," "low mental load," "serendipity," "creative enjoyment," and "exploration of variations."

\begin{figure}[h!]
\centering
\includegraphics[width=0.9\textwidth]{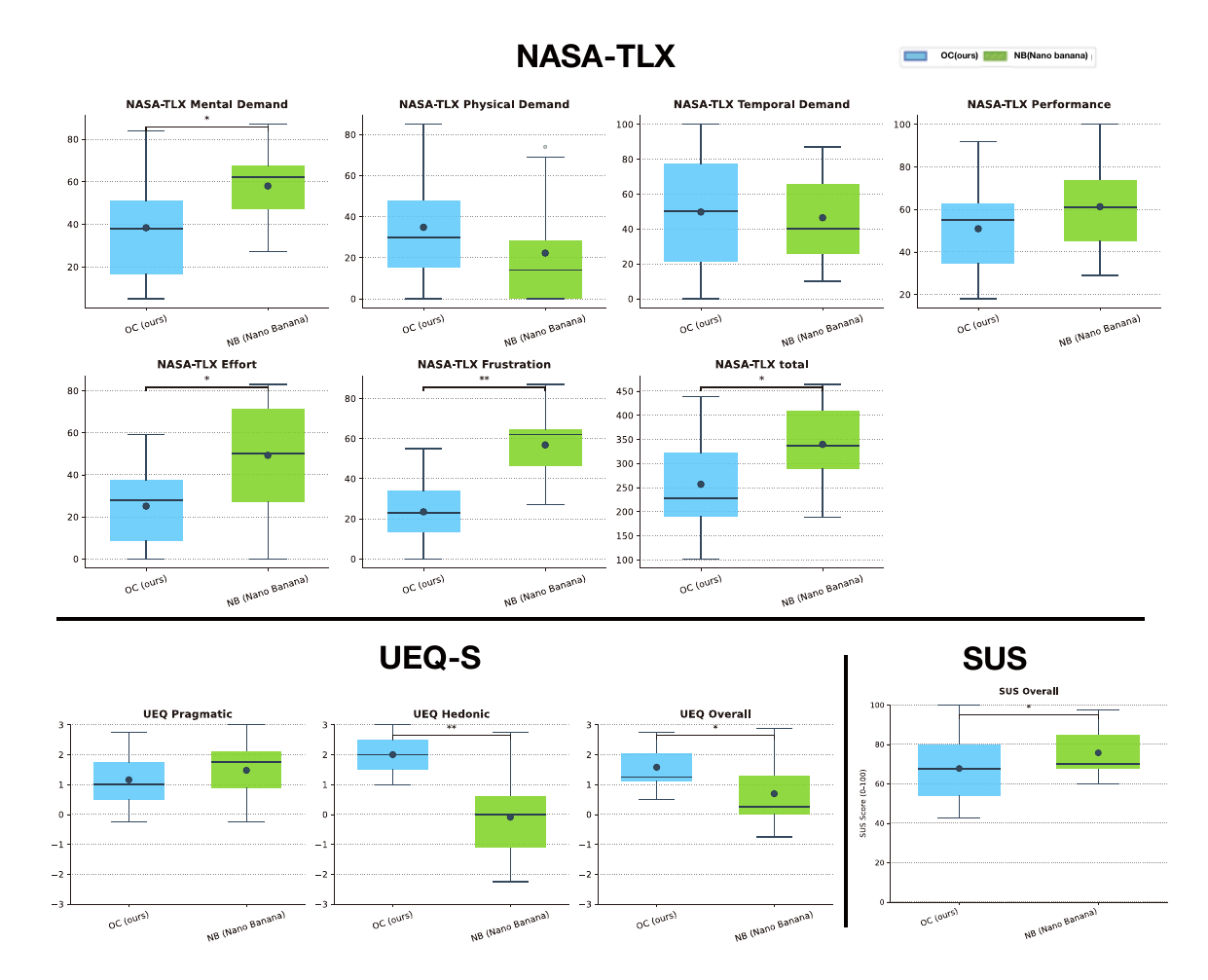}
\caption{\textbf{Comparative evaluation results using standard metrics.} Scores for (a) UEQ (User Experience Questionnaire), (b) NASA-TLX (subjective workload), and (c) SUS (System Usability Scale). In the box plots, the center line represents the median, the dot represents the mean, the box indicates the interquartile range (IQR), and the whiskers represent the range within 1.5 × IQR. (OC: Proposed System [OnomaCompass], NB: Conventional Method [Nano Banana], * p < .05, ** p < .01)}
\end{figure}
\begin{figure}[h!]
\centering
\includegraphics[width=0.9\textwidth]{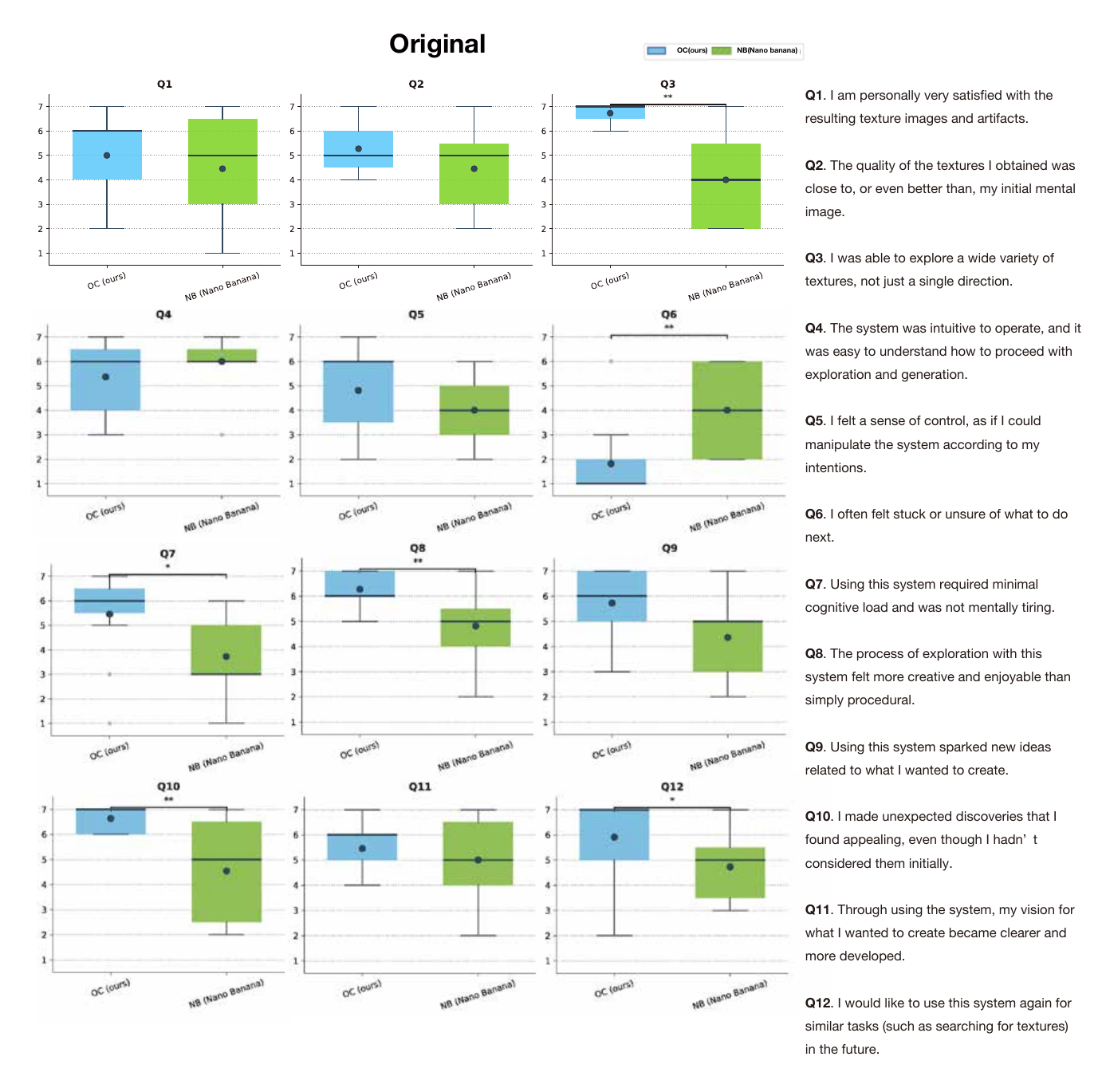}
\caption{\textbf{Evaluation results from the original questionnaire (Q1–Q12).} Response distributions for 12 items (7-point Likert scale) regarding creativity support and exploration experience. OC (Proposed System) received significantly higher ratings, particularly in diverse exploration (Q3), avoidance of getting stuck (Q6), creative process (Q8), and unexpected discoveries (Q10). (* p < .05, ** p < .01)}
\end{figure}

\subsection{Qualitative evaluation of OnomaCompass} \label{sec:qualitative}
\subsubsection{Center of value: Choice-based exploration and discovery}
Participants valued OnomaCompass primarily as an environment for browsing and discovery, especially when they did not have a clearly verbalized target texture in mind. Several contrasted it with prompt-based generation by emphasizing the presence of examples: “With [Nano Banana], there were no examples, so I had to decide many things by myself. With this system, there were many examples aligned with words, so I could choose while looking, which felt easier.”
Participants described their exploration as “clicking onomatopoeia and browsing images to see what happens,” akin to “choosing from a huge wallpaper catalog.” This choice-based interaction lowered the barrier to taking the first step and enabled serendipitous encounters beyond their prior vocabulary and knowledge.

\subsubsection{Onomatopoeia as a cue for reinterpretation and meaning expansion}
Beyond “finding textures,” participants highlighted moments where an onomatopoeia–image pairing reshaped how they understood the expression itself. One participant described a discovery with \textit{shakishaki} (“crisp/crunchy”): while they typically associated the word with the \emph{sound} of biting vegetables, seeing an image that concretely depicted a bitten vegetable surface made the association feel newly grounded—“I had never imagined that visual scene from the word until I saw it.” Another participant commented on \textit{zowazowa} (“creepy-crawly/tingly”), noting that the selected image did not evoke the word in isolation, yet the pairing felt convincing—“I wouldn’t think of \textit{zowazowa} from that image alone, but once paired, it clearly made sense.”
These accounts suggest that onomatopoeia in OnomaCompass functioned less as a precise label to be “matched” and more as a cue that invites reinterpretation and expands the user’s descriptive space.

\subsubsection{Effectiveness as a thinking aid through branching exploration}
OnomaCompass functioned as a thinking aid particularly in early ideation, where goals were still forming. Participants reported that it “compensated for limited vocabulary,” and that the visual map enabled them to proceed by comparison rather than by committing to a single verbal specification. One participant contrasted the cognitive feel of the two systems: “In chat, I feel like I have to reply once the AI responds. With this system, it feels like floating calmly in a big ocean with a swim ring.”
Another participant noted that exploring the texture-image space was efficient because “visual information is easier to judge at a glance,” and that it supported a different kind of ideation: “It made me think ‘combining A and B might be interesting!’”

\subsubsection{Emergent loop and combinatorial ideation for the “in-between”}
Participants particularly valued the interpolation-and-reembedding loop as a creative act rather than mere retrieval. One participant compared it to combinatorial play: “Frame interpolation was fun—it wasn’t just selecting; it had a creative element. Like breeding in Dragon Quest Monsters, exploring what emerges and filling a compendium.”
Notably, participants pointed out that linguistic “in-betweens” are difficult to articulate directly: “It’s hard to name the onomatopoeia between ‘fluffy’ and ‘jagged.’ But interpolating the images feels possible; then seeing what onomatopoeia the AI assigns back is interesting.” This suggests that the system’s value lies in enabling users to explore the conceptual in-between of materialities through a bidirectional loop between images and words.

\subsubsection{Usability challenges}
Conversely, numerous usability issues undermined the system’s value. The most prominent concerns related to interaction in the three-dimensional space: “the 3D operations were difficult without familiarity,” “it demanded 3D manipulation for what felt like planar exploration,” and “it differed from the 3D tools I usually use.” Issues of interaction reliability were also reported, such as “images overlapped and unintended items were selected.” Several participants requested more legible access patterns (e.g., 2D/2.5D navigation, list/search for onomatopoeia, and clearer explanations of what the spatial layout encodes).

\subsection{Qualitative evaluation of the baseline (Nano Banana)}
\subsubsection{Value in controllability and output quality}
The primary strength of Nano Banana was its controllability when participants could verbalize a target clearly. Many praised its ability to specify details (“fine-grained control over which parts used which texture”) and its compliance (“it expressed what was requested faithfully”). Participants also valued the realism and overall image quality.

\subsubsection{Enjoyment in dialogue and “off-the-wall” prompting}
Participants also enjoyed playful interactions such as intentionally unusual prompts and appreciating the AI’s surprising additions (e.g., “When the vase was turned into pudding, a spoon was included”). Some explored novelty by giving words far from material descriptions, expecting “something extra” beyond their own imagination.

\subsubsection{Verbalization barrier, lack of examples, and linear exploration}
However, participants repeatedly identified the burden of verbalization as the central challenge. They noted that texture generation and editing required careful prompt writing, and that “it is difficult to imagine the result until it is generated,” creating a gap between intent and outcome. Several participants also stated that NB provides few concrete examples at the outset, making it hard to decide a direction: “There were no examples, so I had to decide many things by myself.”
Some further characterized the chat workflow as structurally linear: “Because it’s not a format where you pick from multiple options, it becomes one storyline—you proceed step-by-step from the first attempt.” This linearity sometimes led participants to rely on familiar concepts when they got stuck, and some felt that prompt-based generation “can only express what is already in my head,” making it difficult to reach unknown textures beyond their vocabulary.

\section{Discussion}
\subsection{Divergent and convergent thinking: Two models of creativity support}
The interviews strongly suggest that OnomaCompass and prompt-based image-generation AI are complementary, supporting different phases of the creative process. This contrast is aptly explained by Guilford’s concepts of divergent and convergent thinking \cite{Guilford1967}. OnomaCompass was particularly effective during divergent thinking, which expands ideas from a state without a clear goal. In contrast to the image-generation AI, for which “deciding what one likes at the outset was difficult,” participants described OnomaCompass as offering “choices” and “little task burden.” Conversely, image-generation AI excelled during convergent thinking, where a specific goal is progressively realized with precision.

\subsection{Linear dialogue vs. branching exploration in open-ended ideation}
Our qualitative data suggest a structural contrast between prompt-based chat generation and map-based exploration. Participants described NB as requiring continual verbal commitments and unfolding along a single trajectory (“one storyline”), whereas OnomaCompass supported branching exploration by exposing many alternatives simultaneously and enabling reversible, low-cost choices.
This difference is particularly relevant for open-ended ideation, where the target texture is often underspecified and preferences emerge through comparison rather than direct instruction. The “floating calmly” metaphor further suggests that the turn-taking nature of chat can introduce a subtle pressure to respond, while a browsing-oriented interface can better sustain an unhurried exploratory mindset. However, participants sometimes treated the map as a mere collection of items because they could not see what distances and positions meant. This highlights a key design requirement: latent-space interfaces must make the mapping understandable and provide clear ways to act on it.

\subsection{A tool for constructing a “mental map”}
Our findings can be interpreted through differences in how novices and experienced practitioners navigate material concepts. Through repeated exposure and practice, individuals may accumulate a personal repertoire of material references—linking felt impressions, visual examples, and descriptive terms—which can function as an internal “mental map” for specifying intent. Prompt-based image generation can be effective when users already know how to specify their intent within this map. In contrast, our participants—who had limited prior vocabulary for materiality—often benefited from OnomaCompass as a map-construction aid: the linked texture and onomatopoeia maps provided external structure for sensemaking \cite{Pirolli2005}, supporting comparison, naming, and incremental refinement of expectations. In this view, the value of OnomaCompass may lie less in efficiently reaching a single target texture than in providing cognitive scaffolding \cite{wood1976role} that helps users differentiate material impressions and build a reusable set of descriptors (including sound-symbolic ones) for future ideation.

\subsection{Onomatopoeia as an ambiguous but productive bridge}
Our qualitative data also suggest a nuanced role of onomatopoeia: they are not simply labels that map to a single visual texture, but cues that can activate multiple facets of sensory meaning (sound, touch, and situational imagery). Importantly, participants sometimes reported convincing pairings even when the image would not have prompted the word by itself, implying that the value lies in reinterpretation rather than accurate “translation.” This points to a design opportunity: instead of treating ambiguity as noise, interfaces can operationalize sound-symbolic expressions as prompts for meaning-making—helping users articulate and refine material expectations through exposure to grounded examples.

\subsection{The value of the exploration experience and emergent play}
The most important finding is that participants found value in the exploration experience itself. The loop connecting images, videos, and onomatopoeia fostered new discoveries and provided an experience that transcended a mere search tool by actively stimulating creativity. One participant likened the process of combining two materialities to create something new to a popular game series: “In Dragon Quest Monsters, combining species to see what emerges is exciting, similar to the fun of filling a compendium.” This metaphor suggests the system functioned as a space for emergent play, where users actively combine elements to discover unknown outcomes. This constitutes an example of a tool functioning as a cognitive artifact that extends users’ thinking.

\subsection{Limitations and future directions}
The study has limitations. First, the prototype’s usability was immature, potentially preventing participants from fully experiencing the core value of semantic space exploration. Second, the principles governing spatial layout were insufficiently communicated to users. Many reported not understanding “the basis for the mapping,” which may have hindered deeper exploration. Future work should substantially improve usability and examine in greater detail how the principles of spatial layout affect exploratory behavior and creativity. Third, in response to participants’ desire for “both,” designing a hybrid creative environment that seamlessly integrates divergent exploration and convergent generation appears to be a promising direction. Additionally, our dataset includes invented mimetic terms to encourage imagination and avoid reliance on familiar categories. While participants found such terms useful and enjoyable for discovery, we do not claim that the specific associations generalize to broader populations or other languages. Future work should evaluate cross-user agreement and cross-cultural transfer, and compare invented versus conventional ideophones under controlled conditions.

\section{Conclusion}
This paper presented \textit{OnomaCompass}, a web-based AI interface for early-stage materiality ideation that enables users to shuttle between a texture-image space and an onomatopoeia space. Rather than relying on prompt crafting, the system externalizes sound-symbolic cues as navigable structure through (i) dual latent-space visualization with cross-modal highlighting, (ii) curation and texture application previews, and (iii) an emergent loop that generates interpolated videos between textures and re-embeds extracted frames back into both spaces.

A within-subjects study ($N=11$) comparing OnomaCompass with a chat-based prompt-driven image-generation workflow (NB) showed a consistent trade-off. OnomaCompass significantly reduced subjective workload (NASA-TLX overall, mental demand, effort, and frustration) and improved hedonic user experience (UEQ), indicating that spatial shuttling can lower the verbalization burden and make exploratory ideation more enjoyable. Meanwhile, usability (SUS) favored the chat-based baseline, and participants reported concrete interaction issues (e.g., 3D navigation difficulty and selection errors), highlighting that the benefits of exploration-oriented interfaces depend on interaction maturity.

Synthesizing quantitative and qualitative findings, we argue that OnomaCompass is best understood as a divergent-thinking scaffold and a tool for constructing a novice’s “mental map” of materiality, whereas prompt-based generation excels in convergent refinement when users can already specify intent. The system’s value therefore lies not in replacing text-to-image prompting, but in complementing it by making the “in-between” of material concepts explorable and by fostering serendipitous connections through an emergent exploration loop.

Future work will (1) redesign interaction for 2D/2.5D navigation and more reliable selection, (2) better communicate the mapping rationale and uncertainty in cross-modal correspondences, and (3) develop a hybrid workflow that tightly couples divergent latent-space exploration with convergent prompt-based generation and editing. More broadly, this work suggests a direction for co-creative AI interfaces: shifting from one-shot instruction to reciprocal exploration, where users appropriate AI-generated representations as navigable cognitive scaffolding for non-verbal, Kansei-driven domains.
\bibliographystyle{ACM-Reference-Format}

\bibliography{references}

\appendix

\end{document}